\documentclass[nofootinbib,floats,aps,preprint,eqsecnum]{revtex4} 

\newcount\vertou
\ifnum\vertou=0
\def\makeabs#1{\begin{abstract}\vspace*{4mm}#1\end{abstract}\maketitle}
\def\mypreprint#1{\preprint{#1}\vspace*{1cm}}

\def\myaddr#1{\affiliation{\vspace*{3mm}#1\vspace*{1cm}}}
\newcommand{\bim}{\begin{itemize}\vspace*{-4pt}\itemsep-2pt}
\else
\documentclass{ws-procs9x6}
\def\makeabs#1{\maketitle\abstracts{#1}}
\def\mypreprint#1{}
\newcommand{\myaddr}{\address}
\newcommand{\bim}{\begin{itemize}}
\fi

\usepackage{graphicx}
\usepackage{wasysym}
\def\mycenterline#1{\centerline{#1}}

\newcommand{\beqa}{\begin{eqnarray}} 
\newcommand{\eeqa}{\end{eqnarray}} 
\newcommand{\beq}{\begin{equation}} 
\newcommand{\eeq}{\end{equation}}

\newcommand{\eim}{\end{itemize}}

\def\gorer{~~\Longrightarrow~~}
\def\ssb{\quad\Rightarrow\quad}
\def\sun{\astrosun}

\def\vev#1{\langle #1 \rangle}
\def\ket#1{|#1 \rangle}
\def\Tr{\mbox{\rm Tr}}
\def\lsim{\lesssim}
\def\gsim{\gtrsim}

\def\Mpl{M_{\rm Pl}}
\def\Mgut{M_{\rm GUT}}
\def\Mint{M_{\rm Int}}

\begin{document}  
  
\mypreprint{
\vbox{\hbox{hep-ph/0305245}\hbox{May, 2003}}}
  
\title{TASI 2002 lectures on  neutrinos}  
  
\author{Yuval Grossman}  
\myaddr{Department of Physics,  
Technion--Israel Institute of Technology,\\  
       Technion City, 32000 Haifa, Israel}
  
\makeabs{
We present a pedagogical review of neutrino physics. In the first
lecture we describe the theoretical motivation for neutrino masses,
and explain how neutrino flavor oscillation experiments can probe
neutrino masses.  In the second lecture we review the experimental
data, and show that it is best explained if neutrinos are massive.  In
the third lecture we explain what are the theoretical implications of
the data, in particular, what are the challenges they impose on models
of physics beyond the SM.  We give examples of theoretical models that
cope with some of these challenges.}

\section{Introduction}  
The success of the Standard Model (SM) can be seen as a proof that it
is an effective low energy description of Nature. We are therefore
interested in probing the more fundamental theory. One way to go is to
search for new particles that can be produced in yet unreached
energies. Another way to look for new physics is to search for indirect
effects of heavy unknown particles. In this set of lectures we explain how
neutrino physics is used to probe such indirect signals of physics
beyond the SM.

In the SM the neutrinos are exactly massless. This prediction,
however, is rather specific to the SM. In almost all of the SM
extensions the neutrinos are massive and they mix. The search for
neutrino flavor oscillation, a phenomenon which is possible only for
massive neutrinos, is a search for new physics beyond the SM. The
recent experimental indications for neutrino oscillations are
indirect evidences for new physics, most likely, at distances much
shorter than the weak scale.

In the first lecture the basic mechanisms for generating neutrino
masses are described and the ingredients of the SM that ensure
massless neutrinos are explained. Then, the neutrino oscillation
formalism is developed. In the second lecture the current experimental
situation is summarized. In particular, we describe the 
oscillation signals observed by solar neutrino experiments,
atmospheric neutrino experiments and long baseline terrestrial
neutrino experiments.  Each of these results separately can be
accounted for by a rather simple modification to the SM.  Trying to
accommodate all of them simultaneously, however, is not trivial. In
the third lecture we explain what are the theoretical challenges in
trying to combine all the experimental indications for neutrino
masses, and give several examples of models that cope with some of
these challenges.

These lecture notes are aimed to provide an introduction to the topic
of neutrino physics.  They are not meant to be a review. Therefore,
many details are not given and many references are omitted. There are
many textbooks \cite{books} and reviews \cite{review,astro-rev,cy}
about neutrinos. There is also a lot of information about neutrinos on
the web \cite{nu-on-web,lblnews}.  All these sources provide more
detailed discussions with complete set of references on the topics
covered in these lectures. Moreover, they also cover
many subjects that are not mentioned here.

In preparing the lectures I used mainly the recent review by
Gonzalez-Garcia and Nir \cite{cy}. This review is a very good starting
point to anyone who wants to learn more about neutrino physics.
 
\section{Neutrino masses}  

\subsection{Fermion masses}  
In general, there are two possible mass terms for fermions: Dirac and
Majorana mass terms. All fermions can have Dirac mass terms, but only
neutral fermions can have Majorana mass terms. Indeed, all the massive
fermions in the SM, the quarks and charged leptons, have Dirac mass
terms. The neutrinos, however, while massless in the SM, can have both
Dirac and Majorana mass terms.

Dirac masses couple left and right handed fields
\beq \label{dirmass}
m_D \overline{\psi_L} \psi_R + h.c.,
\eeq
where $m_D$ is the Dirac mass and $\psi_L$ and $\psi_R$ are 
left and right handed Weyl spinor fields, respectively. 
Note the following points regarding eq. (\ref{dirmass}):
\bim
\item
Consider a theory with one or more exact $U(1)$ symmetries.  The
charges of $\overline{\psi_L}$ and $\psi_R$ under these symmetries must
be opposite.  In particular, the two fields can carry electric charge
as long as $Q(\psi_L)=Q(\psi_R)$.
\item
Since $\psi_L$ and $\psi_R$ are different fields, there are four
degrees of freedom with the same mass, $m_D$.
\item
When there are several fields with the same quantum numbers, we define
the Dirac mass matrix, $(m_D)_{ij}$
\beq
(m_D)_{ij} (\overline{\psi_L})_i (\psi_R)_j + h.c.,
\eeq
where $i(j)$ runs from one to the number of left (right) handed fields
with the same quantum numbers. In the SM, the fermion fields are present
in three copies, and the Dirac mass matrices are $3 \times 3$
matrices. In general, however, $m_D$ does not have to be a square
matrix.
\eim

A Majorana mass couples 
a left handed or a right handed field to itself. Consider
$\psi_R$, a SM singlet right handed field.  Its Majorana mass term is
\beq \label{majmass}
m_M \overline{\psi_R^c} \,\psi_R, \qquad \psi^c = C \,\overline{\psi}^T,
\eeq
where $m_M$ is the Majorana mass and $C$ is the charge
conjugation matrix \cite{peskin}. A similar expression holds for left
handed fields.
Note the following points regarding eq. (\ref{majmass}):
\bim
\item
Since only one Weyl fermion field is needed in order to generate a
Majorana mass term, there are only two degrees of freedom that have
the same mass, $m_M$.
\item
When there are several neutral fields, $m_M$ is promoted to be a
Majorana mass matrix
\beq
(m_M)_{ij} (\overline{\psi_R^c})_i (\psi_R)_j.
\eeq
Here, $i$ and $j$ runs from one to the number of neutral fields. A
Majorana mass matrix is symmetric.
\item
The additive quantum numbers of $\overline{\psi_R^c}$ and $\psi_R$ are
the same. Thus, a fermion field can have a Majorana mass only if it is
neutral under all unbroken local and global $U(1)$ symmetries. In
particular, fields that carry electric charges cannot acquire Majorana
masses.
\item
We usually work with theories where only local symmetries are
imposed. In such theories global symmetries can only be accidental. A
Majorana mass term for a fermion field $\psi$ breaks all the global
$U(1)$ symmetries under which the fermion is charged.
\eim

\subsection{Neutrino masses in the SM}  

The SM is a renormalizable four dimensional
quantum field theory where
\bim
\item
The gauge group is $SU(3)_C \times SU(2)_L \times U(1)_Y$.
\item
There are three generations of fermions
\beqa
Q_L(3,2)_{+1/6}, && \quad
U_R(3,1)_{+2/3}, \qquad 
D_R(3,1)_{-1/3}, \nonumber \\
L_L(1,2)_{-1/2}, && \quad
E_R(1,1)_{-1}\,.
\eeqa
We use the $(q_C,q_L)_{q_Y}$ notation, such that $q_C$ is the
representation under $SU(3)_C$,  $q_L$ is the
representation under $SU(2)_L$ and  $q_Y$ is the
$U(1)_Y$ charge.  
\item
The vev of the scalar Higgs field, $H(1,2)_{+1/2}$, leads to the
Spontaneous Symmetry Breaking (SSB)
\beq
SU(2)_L \times U(1)_Y \to U(1)_{EM}\,.
\eeq
\eim

The SM has four accidental global symmetries: baryon number ($B$) and
lepton flavor numbers ($L_e$, $L_\mu$ and $L_\tau$). It is also
convenient to define total lepton number as $L=L_e+L_\mu+L_\tau$,
which is also conserved in the SM. An accidental symmetry is a
symmetry that is not imposed on the action. It is there only due to
the field content of the theory and by the requirement of
renormalizability. If we would not require the theory to be
renormalizable, accidental symmetries would not be present. For
example, in the SM dimension five and six operators break lepton and
baryon numbers.
 
The SM implies that the neutrinos are exactly massless. There are
several ingredients that combine to ensure it:
\bim
\item 
The SM does not include fields that are singlets under the gauge
group, $N_R(1,1)_0$. This implies that there are no Dirac mass terms
$\vev{\tilde H} \overline{\nu_L} \nu_R$.\footnote{Our notation is such
that we use capital letters to denote fields before SSB and lowercase
Greek or Roman letters to denote the fields after SSB. While such
differentiation is meaningful only to field that are charged under the
SM gauge group, we use this convention also for SM singlets.}  (We
define $\tilde H \equiv i \tau_2 H^*$.)
\item 
There are no scalar triplets, $\Delta(1,3)_1$, in the SM.
Therefore, Majorana mass terms of the 
form $\vev{\Delta} \overline{\nu_L^c} \nu_L$ cannot be written.
\item 
The SM is renormalizable. This implies that no dimension five Majorana
mass terms of the form $\vev{H}\vev{H} \overline{\nu_L^c} \nu_L$ are
possible.
\item 
$U(1)_{B-L}$ is an accidental non-anomalous global symmetry of the
SM. Thus, quantum corrections cannot generate Majorana $\vev{H}\vev{H}
\overline{\nu_L^c} \nu_L$ mass terms.\footnote{Note that $U(1)_{B-L}$
has gravitational anomalies. Namely, once gravity effects are included
the symmetry is broken and neutrino masses can be generated. The SM,
however, does not include gravity. What we learn is that it is likely
that the neutrinos are massive in any SM extension that include
gravity, in particular, in the real world.}
\eim

Both the neutrinos and the $SU(3)_C \times U(1)_{EM}$
gauge bosons are massless in the SM. There is,
however, a fundamental difference between these two cases. The symmetry
that protects the neutrinos from acquiring masses is lepton number,
which is an accidental symmetry. Namely, one does not 
impose it on the SM.  In contrast, the photon and gluon
are massless due to local gauge invariance, which is 
a symmetry that one imposes on the theory. This difference is
significant. An imposed symmetry is exact also when one considers
possible new physics at very short distances. Accidental symmetries,
however, are likely to be broken by new heavy fields.

One can understand the result that the neutrinos are massless in the
SM in simple terms as follows. For a massive particle one can always
find a reference frame where the particle is left handed and another
reference frame where it is right handed. Thus, in order to have
massive neutrinos, the SM left handed neutrino fields should couple to
right handed fields. There are two options for such couplings.  First,
the SM left handed neutrinos can couple to right handed SM singlet
fermions. This is not allowed in the SM just because there are no such
fields in the SM. The second option is to have couplings between the
left handed neutrinos and the right handed anti-neutrinos. Such 
couplings break lepton number. The fact that lepton number is an
accidental symmetry of the SM forbids also this possibility.

\subsection{Neutrino masses beyond the SM}  

Once the SM is embedded into a more fundamental theory, some of its
properties mentioned above are lost.  Therefore, in many SM extensions
the neutrinos are massive.  Neutrino masses can be generated by adding
light or heavy fields to the SM. (By light we refer to fields that
have weak scale or smaller masses, while heavy means much above the
weak scale.)  When light fields are added the resulting neutrino
masses are generally too large and some mechanism is needed in order
to suppress them. New heavy fields, on the contrary, generate very
small masses to the neutrinos. We are now going to elaborate on these
two possibilities.
 
There are two kinds of light fields that can be used to generate
neutrino masses. First, right
handed neutrino fields, $N_R(1,1)_0$, couple to the SM left handed
lepton fields via the Yukawa couplings and generate Dirac masses for
the neutrinos
\beq
y_N \tilde H \overline{L_L} N_R + h.c. \ssb m_D \overline{\nu_L} \nu_R,
\eeq
where $y_N$ is a dimensionless Yukawa coupling and the symbol
``$\Rightarrow$'' indicates electroweak SSB. 
This mechanism offers no explanation
why the neutrinos are so light and why the right
handed neutrinos do not acquire large Majorana masses. Another way to
generate neutrino masses is to add to the SM a scalar triplet,
$\Delta(1,3)_1$. Assuming that the neutral component of 
this triplet acquires a vev, a Majorana 
mass term is generated
\beq
\lambda_N \Delta  \overline{L^c_L} L_L \ssb m_N \overline{\nu_L^c} \nu_L,
\eeq
where $\lambda_N$ is a dimensionless coupling. In that case it remains
to be explained why the triplet vev is much smaller than the SM Higgs
doublet vev. Note that the triplet vev is required to be small not
only to ensure light neutrinos but also from electroweak precision
data, in particular, from the measurement of the $\rho$ parameter
\cite{pdg}.

The other way to generate neutrino masses is to add new heavy fields
to the SM. In that case the low energy theory is the SM, but the full
high energy theory includes new fields. Using an effective field
theory approach, the effects of these heavy fields are described by
adding non-renormalizable terms to the SM action. All such terms are
suppressed by powers of the small parameter $v/M$. Here $v$ is the SM
Higgs vev, which characterizes the weak scale, and $M$ is the unknown
scale of the new physics, which can be, for example, the Planck scale
or the GUT scale.  Neutrino masses are generated by the following
dimension five operator
\beq \label{genmass}
{\lambda \over M} HH \overline{L_L^c} L_L \ssb
m_\nu = \lambda {v^2 \over M},
\eeq
where $\lambda$ is a dimensionless coupling. 

Note the following points regarding eq. (\ref{genmass}):
\bim
\item
Taking the SM to be an effective field
theory implies $m_\nu \ne 0$. Since we know that
new physics must exist at, or below, the Planck scale, it is 
rather likely that the neutrinos are massive.
\item
$m_\nu$ is small since it arises from non-renormalizable terms.
\item
Neutrino masses probe the high energy physics.
\item
The neutrino acquires Majorana mass. Thus, total lepton number is
broken by two units.
\item
The situation can be generalized to the case of several generations.
Then, generically, both total lepton number and family lepton numbers
are broken and lepton mixing and CP violation are expected.
\eim

\begin{figure}[t]
\centerline{\includegraphics[height=66mm]{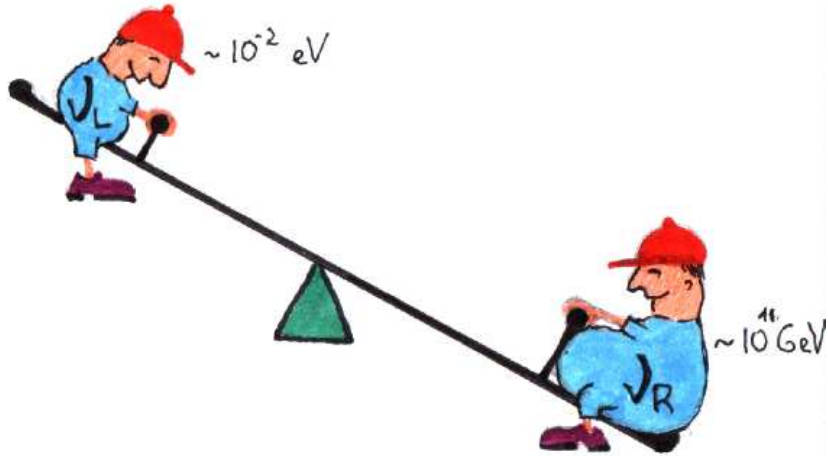}~~~~~~}
\caption[y]{The seesaw mechanism acquire its name from analogy with
a children seesaw. The lower one child takes its arm of the
seesaw, the higher the other child goes. For neutrinos, the larger
$M_N$ is, the lighter the neutrino becomes.  (I Thank Oleg Khasanov
for the figure.) \label{see-saw-fig}}
\end{figure}

A famous example of a realization of this effective field theory
description is the seesaw mechanism, see fig. \ref{see-saw-fig}.
(Historically, it was invented within $SO(10)$ GUT theory, and later
implemented in many other models.) 
Consider one generation SM with an additional fermion singlet
$N_R(1,1)_0$. The following two terms in the Lagrangian are relevant
for neutrino masses
\beq
{1 \over 2} M_N \overline{N_R^c} N_R + Y_\nu H \overline{L_L} N_R  + h.c.,
\eeq 
where $M_N \gg v$ is the Majorana mass of the right handed neutrino.
After electroweak symmetry breaking the second term leads to a Dirac
mass of the neutrino. In the $(\nu_L,\nu_R^c)$ basis the neutrino mass
matrix is
\beq
m_\nu=\pmatrix{0 & m_D \cr m_D & M_N}\,,
\eeq
where $m_D \equiv Y_\nu v$.
Using 
\beq
\Tr(m_\nu)=M_N\,, \qquad
|\det(m_\nu)|=m_D^2\,,
\eeq
to first order in $v/M_N$ we find
\beq
m_{\nu_R}=M_N \qquad m_{\nu_L}={m_D^2 \over M_N}\,. 
\eeq
Comparing this result with eq.~(\ref{genmass}) one identifies the new
physics scale $M$ with $M_N$ and the new physics coupling $\lambda$
with $Y_\nu^2$.  We learn that the seesaw mechanism is indeed a
realization of the effective field theory approach for neutrino
masses.

\subsection{Neutrino mixing}

Massive neutrinos generally mix. The neutrino mass terms break the
accidental family lepton number symmetries. (Total lepton number is
also violated if the neutrinos have Majorana masses.) This phenomenon
is very similar to quark mixing in the SM.  It is therefore
instructive to describe both lepton and quark mixing in parallel.

For quarks, the Cabibbo-Kobayashi-Maskawa (CKM)
matrix, $V$, corresponds to  non-diagonal charged current interactions
between quark mass eigenstates
\beq
{g \over \sqrt{2}} (\overline{u_{L}})_i V_{ij} \gamma^\mu (d_{L})_j
W^+_\mu\,, \qquad i=u,c,t\,, \quad j=d,s,b\,.
\eeq
For leptons, it is common to use two different bases.
The  flavor basis is defined to be the one where the charged lepton
mass matrix and the $W$ interactions are diagonal.
In the mass basis both the charged lepton and the neutrino mass
matrices are diagonal, but the $W$ interaction is not. In that basis
the leptonic mixing
matrix $U$ is the analogue of the CKM matrix.\footnote{Recently, in
many papers the matrix $U$ is called the Maki-Nakagawa-Sakata (MNS)
matrix, $U_{\rm MNS}$, or the Pontecorvo-Maki-Nakagawa-Sakata (PMNS)
matrix, $U_{\rm PMNS}$, since these authors were the first to discuss
leptonic flavor mixing.  Others call it the CKM or KM analogue for the
lepton sector since Kobayashi and Maskawa were the first to discuss CP
violation from such matrices. Here we adopt the notation of
\cite{cy} and call $U$ the leptonic mixing matrix. The fact that
the CKM matrix is denoted by $V$ helps in avoiding confusion.}
Namely, it shows up in the charged current interactions
\beq
{g \over \sqrt{2}}\, \overline {\ell_L} U_{\ell i} \gamma^\mu (\nu_{L})_i
W^-_\mu\,, \qquad \ell=e,\mu,\tau\,, \quad i=1,2,3\,.
\eeq

When working in the mass basis, the formalism of quark and
lepton flavor mixings are very similar. The difference between these two
phenomena arises due to the way neutrino experiments are done. While quarks
and charged leptons are identified as mass eigenstates, neutrinos
are identified as flavor eigenstates.  Indeed, there are two
equivalent ways to think about fermion mixing.  For quarks, mixing is best
understood as the fact that the $W$ interaction is not diagonal in the
mass basis.  For leptons, we  usually refer to the rotation between
the neutrino mass and flavor bases as neutrino mixing.

Note that the indices of the two matrices are of different order. In
$V$ the first index corresponds to the up type component of the
doublet and the second one to the down type. In $U$ it is the other
way around.

In general, $U$ is described by three mixing angles and one Dirac CP
violating phase. If the neutrinos are Majorana particles there are in
addition two more CP violating phases called Majorana phases. The
Majorana phases, however, do not affect neutrino flavor
oscillations, and we do not discuss them any further.

SM singlet states can mix with the three SM neutrinos. When the masses
of the singlet states are small they can participate in neutrino
oscillation. Then, the singlet states are usually called sterile
neutrinos while the standard model neutrinos are called active
neutrinos.\footnote{Oscillation occurs only between left handed
states, that is, both active and sterile states are left handed. One
should not get confused by the fact that the sterile states are left
handed; the identification of left handed states as SM doublets and
right handed states as SM singlets does not have to be true in
theories that extend the SM.}  The mixings between all neutrinos are
taken into account by enlarging the mixing matrix $U$ to a $3\times
(n+3)$ matrix, where $n$ is the number of sterile states.

When the SM singlet states are very heavy (with mass $M \gg v$), they
cannot participate in neutrino oscillations. Their couplings to the
light neutrino is important since it gives them their masses. Thus, the
heavy states are usually referred to as right handed neutrinos. The
mixings relevant for neutrino oscillations is described by the matrix
$U$ that refers only to the $3\times 3$ sub-matrix that corresponds to
the mixing between the mostly active states. The presence of the heavy
states affects the theory since in that case the matrix $U$ does not
describe the full mixing and, therefore, it is not unitary. This
deviation from unitarity, however, is very small, of order $v/M$, and
it is usually neglected.

\section{Methods for probing neutrino masses}
The conclusion of the last section is that it is very likely that the
neutrinos have very small masses and that they mix. There are several
ways to probe neutrino masses and mixing angles.  Discussions about
astrophysical and cosmological probes of neutrino masses can be found,
for example, in \cite{books,astro-rev}.  Here we briefly discuss
kinematic tests for neutrino masses and neutrinoless double beta
decay.  Then we develop the formalism of neutrino oscillation, which
is the most promising way to probe the neutrino masses and mixing
angles.

\subsection{Kinematic tests}
In decays that produce neutrinos the decay spectra are sensitive
to neutrino masses.  For example, in $\pi
\to \mu \nu$ the muon momentum is fixed (up to tiny width effects), and
it is determined only by the masses of the pion, the muon and the
neutrino. To first order in $m_\nu^2/m_\pi^2$, the muon momentum in the
pion rest frame is given by
\beq
|\vec p\,| = {1 \over 2m_\pi}\left(m_\pi^2 - m_\mu^2 - {m_\pi^2 + m_\mu^2
\over m_\pi^2 - m_\mu^2}\, m_\nu^2 \right)\,.
\eeq
Since the correction to the massless neutrino 
limit is proportional to $m_\nu^2$, the kinematic tests are
not very sensitive to small neutrino masses. The current best bounds
obtained using kinematic tests are \cite{pdg} 
\beqa
m_\nu < 18.2~{\rm MeV} && \quad {\rm from~~} \tau \to 5\pi +\nu\,, \nonumber \\
m_\nu < 190~{\rm KeV} && \quad   {\rm from~~} \pi \to \mu \nu\,, \nonumber \\
m_\nu < 2.2~{\rm eV} && \quad   {\rm from~~} ^3{\rm H} \to ^3\!\!{\rm He} + e +
\overline\nu\,.
\eeqa
The sensitivities of neutrino oscillation experiments are much better
than these bounds.  Note that while oscillation experiments are
sensitive to the neutrino mass-squared differences (see below) the
kinematic tests are sensitive to the neutrino masses themselves.

\subsection{Neutrinoless double $\beta$ decay}
Neutrino Majorana masses violate lepton number by two
units. Therefore, if neutrinos have Majorana masses we expect there
there are also $\Delta L=2$ processes.  The smallness of the neutrino
masses indicates that such processes have very small rates. Therefore,
the only practical way to look for $\Delta L=2$ processes is in places
where the lepton number conserving ones are forbidden or highly
suppressed. Neutrinoless double $\beta$ decay where the single beta
decay is forbidden is such a process.  An example for such processes
is
\beq
 ^{32}_{76}{\rm Ge} \to\, ^{34}_{76}\,{\rm Se} + 2e^-\,.
\eeq
The only physical background to neutrinoless double $\beta$ decay is
from double $\beta$ decay with two neutrinos.

Note the following points:
\bim
\item
Since  neutrinoless double
$\beta$ decay  is a $\Delta L=2$ process, it is sensitive only to Majorana
neutrino masses. Dirac masses conserve lepton number and do not
contribute to this decay.
\item 
The neutrinoless double $\beta$ decay rate is related to the neutrino
mass-squared. It is also proportional to some nuclear matrix
elements. Those matrix elements introduce theoretical uncertainties in
extracting the neutrino mass from the signal, or in deriving a bound
if no signal is seen.
\item
Neutrinoless double $\beta$ decay is sensitive to any $\Delta L=2$
operator, not only to the neutrino Majorana masses.
Thus, the relation between the Neutrinoless double $\beta$ decay rate
and the neutrino mass is model dependent.
\item
The best bound derived from neutrinoless double $\beta$ decay is
$m_\nu < 0.34 ~{\rm eV}$ \cite{heidel}. 
\eim

\subsection{Neutrino vacuum oscillation}

We now turn to the derivation of the neutrino oscillation formalism.
We use several assumptions in this derivation.  We
emphasize, however, that in all practical situations it is correct to
use these assumptions and that more sophisticated derivations give the
same results (see, for example, \cite{Grossman-Lipkin}).

In an ideal neutrino oscillation experiment, a neutrino beam is
generated with known flavor and energy spectrum.  The flavor and the
energy spectrum is then measured at some distance away.  If the flavor
composition of the beam changed during the propagation it is a signal
of neutrino oscillation, which indicates neutrino masses and mixings.

The flavor of the neutrino is identified via its charged current interaction. 
It is convenient to express a flavor state in terms of mass states
\beq
\ket{\nu_\alpha} = \sum_i U^*_{\alpha i}\ket{\nu_i},
\eeq
where Greek indices runs over the flavor states ($\alpha=e,\mu,\tau$)
and Roman ones over the mass states ($i=1,2,3)$.  An initially
produced flavor state $\ket{\nu_\alpha}$ evolves in time according to
\beq
\ket{\nu_\alpha(t)}=\sum_i e^{-i E_i t} U^*_{\alpha i}\ket{\nu_i},
\eeq
where we assume that all the components in the neutrino wave packet have the
same momenta ($p_i=p$). For ultra relativistic
neutrinos we can use the following approximation
\beq
E_i \approx p + {m_i^2 \over 2 p}.
\eeq
Then, the probability to observe oscillation between flavor $\alpha$
and $\beta$ is given by
\beq
P_{\alpha \beta}=\left|\vev{\nu_\beta|\nu_\alpha(t)}\right|^2
=\delta_{\alpha \beta}-4 \sum_{i<j} 
U_{\alpha i}U_{\beta i}U_{\alpha j}U_{\beta j}] \sin^2 x_{ij},
\eeq
where in the last step we assumed CP conservation and
\beq
x_{ij}={\Delta m^2_{ij} t \over 2p}.
\eeq
In many cases the three flavor oscillation is well approximated by
considering only two flavors. Then, 
\beq
U=\pmatrix{\cos\theta & \sin \theta \cr -\sin\theta & \cos \theta} ,
\eeq
and for $\alpha \ne \beta$ we obtain
\beq \label{masosc}
P_{\alpha \beta} = \sin^2 2\theta \, \sin^2 x \,.
\eeq
Since the neutrinos are relativistic we usually use $L=t$, $E=p$ and
write
\beq
x = {2 \pi L \over L_{\rm osc}}, \qquad   
L_{\rm osc}= {4 \pi E \over \Delta m^2}\,,
\eeq
such that $L_{\rm osc}$ is the oscillation length. 
The following formula is also convenient
\beq
x\approx 1.27 \left(\Delta m^2 \over {\rm eV}^2\right)
\left(L \over {\rm km}\right)\left({\rm GeV} \over E\right)\,.
\eeq

The oscillation master formula, eq. (\ref{masosc}), teaches us the following:
\bim
\item Longer baseline, $L$, or smaller energy, $E$, is needed in order to probe
smaller $\Delta m^2$. 
\item When $L \ll L_{\rm osc}$ we can approximate
$\sin^2 x \sim x^2$. In that case  the signal is usually too small to
be detected.
\item When $L \gg L_{\rm osc}$ the energy spread of the beam and
decoherence effects cause the oscillations to average out. That is,
in our formula we should take the average value:
$\vev{\sin^2x}=0.5$. In that case only a lower bound on the 
mass-squared difference between the two neutrinos can be obtained.
\eim


\subsection{Matter effects}

When neutrinos travel through matter the oscillation formalism is
modified. While neutrinos can scatter off the medium, in almost all
relevant cases the scattering cross sections are very small, and the
effect of scattering is negligible. The effect we are concentrating on
is that of the forward scattering of the neutrinos. Like photons, when
neutrinos travel inside a medium they acquire effective masses. There
is no energy or momentum exchange between the neutrinos and the
medium. The effect of the medium is that it induces effective masses
for the neutrinos.

First we discuss the case of a medium with constant density.
Moreover, we consider only matter at low temperature, which implies
that it consists of electrons, protons and neutrons. Note that there
are no muons, taus and anti-leptons in that case.  The charged
current interaction between the electron neutrinos and the electrons
in the medium induces an effective potential for the neutrinos
\cite{Wol}
\beq
V_C=\sqrt{2} G_F N_e\approx 7.6\, Y_e 
\left({\rho \over 10^{14} {\rm g/cm^3}}\right) {\rm eV} \,,
\eeq
where $N_e$ ($N_p$, $N_n$) is the number density of electrons
(protons, neutrons) and $Y_e=N_e/(N_p+N_n)$ is the relative
electron number density. To get a feeling for the size of this
potential, note that at the Earth core $\rho \sim 10 \;{\rm g/cm^3}$,
which gives rise to $V_C \sim 10^{-13}\;{\rm eV}$.

As we discuss below, the current data indicates that $m_\nu
\gsim 10^{-3}$ eV, and thus $m_\nu \gg V_C$. This seems to suggest
that matter effects are irrelevant. This is, however, not the case
since the matter effects arise from vector interactions while masses are
scalar operators. The right comparison to make is between $m_\nu^2$
and $E V_C$ where $E$ is the neutrino energy.  Since $E \gg m_\nu$
matter effects are enhanced and can be important.

To explain this enhancement, we consider a uniform, unpolarized medium
at rest. (Discussion about the general case which includes all types
of interactions and arbitrary polarization can be found, for example,
in \cite{bgn}.) In that case the neutrino feels the four-vector
interaction $V_\mu=(V_C,0,0,0)$.  Due to $V_C$ the vacuum dispersion
relation of the neutrino, $p_\mu p^\mu=m^2$, is modified as follows
\beq
(p_\mu-V_\mu)(p^\mu-V^\mu)=m^2 
\gorer E \approx p + V_C + {m^2 \over 2 p}\,,
\eeq
where the approximation holds for ultra relativistic neutrinos.
We learn that the effective 
mass-squared in matter, $m^2_m$, is given by
\beq \label{defA}
m^2_m=m^2 + A, \qquad A\equiv 2 E V_C\,,
\eeq
where we used $p \approx E$. It is the vector nature of the weak
interaction that makes the matter effects practically relevant.

Interactions that are flavor universal only shift the neutrino energy
by a negligibly small amount and do not affect the
oscillation. Non-universal interactions, however, are important since
in their presence the values of the effective masses and mixing angles
are different from their vacuum ones. While the weak neutral current
is flavor universal, the charged current is not.  In normal matter,
where there are electrons but not muons and taus, only electron
neutrinos interact via the charged current with the medium.  The
mixing matrix is modified by this non-universal matter effect such
that the effective squared mass difference and mixing angle are given
by
\beqa \label{masmat}
\Delta m^2_m &=&\sqrt{(\Delta m^2 \cos 2\theta-A)^2 + 
 (\Delta m^2 \sin 2\theta)^2},
\\
\tan 2 \theta_m &=& {\Delta m^2 \sin 2 \theta \over \Delta m^2 \cos 2\theta-A},
\nonumber
\eeqa
where the subindex $m$ stand for 
matter.  The oscillation probability is then obtained from
(\ref{masosc}) by replacing the vacuum masses and mixing angles by
the corresponding parameters in matter
\beq \label{oscmat}
P_{\alpha \beta} = \sin^2 2\theta_m \sin^2 x_m\,, \qquad 
x_m={\Delta m^2_m L \over 2E}\,.
\eeq

The following points are worth mentioning regarding eq. (\ref{masmat}):
\bim
\item
The vacuum result is reproduced for $A=0$, as it should.
\item
Vacuum mixing is needed in order to get mixing in matter.
\item
To first order in $x_m$ the matter effects cancel in the oscillation
probability. To see this note that 
$x\sin 2\theta= x_m \sin 2\theta_m +O(x_m^3)$.
Therefore, when approximating $\sin x_m \sim x_m$ eq. (\ref{oscmat}) 
reduces to the oscillation probability in vacuum, eq. (\ref{masosc}).
\item
For $\Delta m^2 \cos 2 \theta \gg |A|$, the matter effect is a small
perturbation to the vacuum result. 
\item
For $\Delta m^2 \cos 2 \theta \ll |A|$, the neutrino mass is  a small
perturbation to the matter effect. In that case
the oscillations are highly suppressed since the effective mixing
angle is very small.
\item
For $\Delta m^2 \cos 2 \theta = A$, the mixing is maximal, namely it is
on resonance.
\eim

\subsection{Non-uniform density}
When the matter density is not constant there are further
modifications to the oscillation formalism. Density variation results
in changing the effective neutrino masses and their mixing
angles. Then, the  flavor composition of the neutrinos
along their path is a function of the medium density profile.

At any point $r$ on the neutrino path we define the derivative of the
density     
\beq 
A'(r)={dA(r) \over dr}.
\eeq
For constant density, $A'=0$, the flavor conversion probability is
controlled by the effective masses and mixing angles.  For varying
density, $A'\ne 0$, there are extra parameters that affect 
the flavor conversion probability. Most important is 
the adiabatic parameter
\beq
Q(r)={\Delta m^2 \sin^2 2 \theta \over E\, \cos2\theta} \, {A(r) \over
A'(r)}.
\eeq
In the adiabatic limit, $Q \gg 1$, the density variation is slow. In
this case the transition between effective mass eigenstates is highly
suppressed, and the constant density formalism can be applied locally.
In the non-adiabatic limit, $Q <1$, the density variation is
fast. Then, transition between effective mass eigenstates is possible,
and the constant density formalism cannot be used. Both limits can be
of interest in reality.

\begin{figure}[t]
\mycenterline{\includegraphics[height=7cm]{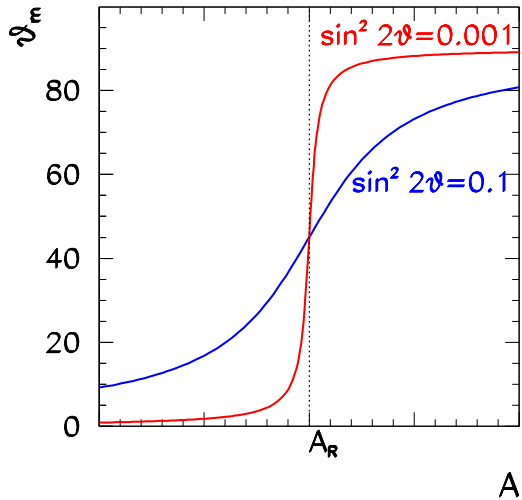}}
\caption[a]{A demonstration of an MSW resonance conversion. $\theta_m$
is the effective mixing angle in matter, $A$ is the effective mass due
to the matter effect defined in (\ref{defA}) and $A_R$ is the
resonance point.  The neutrinos are produced at a region where
$A>A_R$. They propagate as effective mass eigenstates, which is shown
in the figure as solid lines. In the plot the effect is demonstrated
for two different values of the vacuum mixing angles. (The plot is
taken from
\cite{cy}.)\label{fig-mix}}
\end{figure}

The best known example of the effect of the density variation is the
MSW effect \cite{Wol,msw}.  It can account for a very efficient flavor
conversion in the case of small mixing angle. Consider a two generation
model where $m_2>m_1$ with small vacuum mixing angle, $\theta \ll 1$.
That is, in vacuum the muon neutrino is mainly $\nu_2$ and the electron
neutrino is mainly $\nu_1$.  We consider neutrinos that are produced
in the Sun core where the density is large.
In particular, we are interested in the case where at the core of the
Sun, $r=0$, the matter induce effective electron neutrino
mass-squared is much larger than the vacuum mass-squared difference,
$A(0) \gg \Delta m^2$.  Then, at $r=0$ flavor eigenstates are almost
pure effective mass eigenstates
\beq
\theta_m(r=0) \to \pi/2, \qquad
\nu_e(r=0) \approx \nu_2^m, \qquad
\nu_\mu(r=0) \approx \nu_1^m.
\eeq
(By writing $\nu_\alpha \approx \nu_i^m$ we mean that a very small
rotation in flavor space is needed in order to move from the flavor
eigenstate $\nu_\alpha$ to the effective mass eigenstate $\nu_i^m$.)
In particular, the produced electron neutrino is almost a pure
$\nu_2$.  Also at the edge of the Sun,  $r=R_{\sun}$, the flavor
eigenstates are almost pure mass eigenstates
\beq
\theta_m(r=R_{\sun}) \to \theta \ll 1, \qquad
\nu_e(r=R_{\sun}) \approx \nu_1, \qquad
\nu_\mu(r=R_{\sun}) \approx \nu_2.
\eeq
When the adiabatic limit applies, neutrinos propagate in the Sun as
effective mass eigenstates. Since the neutrinos are produced as almost
pure $\nu_2$, they leave the Sun as $\nu_2$. Since in vacuum $\nu_2$
is almost a pure $\nu_\mu$, we learn that there is almost full
conversion from $\nu_e$ to $\nu_\mu$.  This phenomenon, which is
called MSW resonance conversion, is demonstrated in
fig. \ref{fig-mix}.

In the more general case, corrections to the adiabatic limit are taken
into account. In particular, transitions between the effective mass
eigenstates are important in the vicinity of the resonance. 
Such level crossing is described by the Landau-Zener probability, 
$P_{LZ}$.
Then, the flavor conversion probability is given by the Parke
formula \cite{Parke}
\beq
P_{ee} = {1 \over 2} \left[1+(1-2 P_{LZ})\cos 2 \theta_m\,
\cos 2\theta\right]\,,
\eeq
where $\theta_m$ is the effective mixing angle at the production
point.  In this general case, the flavor transition probabilities
become a rather involved function of the neutrino and medium
parameters.  It is this richness that makes the flavor composition of
the solar neutrinos a complicated and not a monotonic function of
their energies.  As it is shown below this is important in order to be
able to explain the solar neutrino data.

\subsection{Neutrino oscillation experiments}
Ideally, neutrino oscillation experiments measure the neutrino flavor
transition probabilities as a function of the neutrino energies and
their path.  In practice, however, there are many experimental
complications.  The basic setup of any neutrino oscillation experiment
is as follows: Neutrinos are produced and then propagate until few of
them are detected at another location, far away from their production
point. Therefore, in order to be able to probe the fundamental physics
parameters we would like to know and control the following:
\bim
\item
The total flux, the flavor composition and the energy spectrum of the
beam.
\item
The distance traveled by the neutrinos and the matter profile that
they passed.
\item
The total
flux, the flavor composition and the energy spectrum of the detected
neutrinos. 
\eim
In reality, it is impossible to know and control all of these
parameters. Therefore, a lot of work is involved in order to get
the maximum out of each experimental setup.

In general we distinguish between two kinds of experiments.
Disappearance experiments search for  reduction in the flux of a
specific neutrino flavor 
\beq
N[\nu_\alpha(L)] < N[\nu_\alpha(0)]\,,
\eeq
where $N[\nu_\alpha(L)]$ is the number of neutrinos of flavor
$\alpha$ in the neutrino beam at distance $L$. This would imply
\beq
P_{\alpha \alpha}(L) < 1 \,.
\eeq
Appearance experiments, on
the other hand, look for enhancement of a flux of a specific flavor
\beq
N[\nu_\beta(L)] > N[\nu_\beta(0)]\,,
\eeq
which would imply
\beq
P_{\alpha \beta}(L) > 0.
\eeq
In particular, in many cases $N[\nu_\beta(0)]=0$. Then, any
observation of neutrinos with flavor $\beta$ is an appearance signal.

Oscillation experiments are usually named after their production
place and mechanism. In the following we discuss solar neutrinos,
atmospheric neutrinos and terrestrial neutrinos.

\section{Solar neutrinos}
Solar neutrinos receive a lot of attention since there are
indications for disappearance of electron neutrinos. These indications
are collectively referred to as the solar neutrino problem or the
solar neutrino anomaly. The best explanation for the solar neutrino
problem is neutrino flavor oscillation.

The use of the words ``problem'' or ``anomaly'' should not worry
us. They are used only to indicate the fact that solar neutrinos pose
a problem to the SM. We think, however, that the SM is an effective
theory of Nature, and thus, the fact that it faces problems is not
surprising. Actually, we expect that neutrinos have masses and that
they mix, and consequently, that the solar neutrino data cannot be
explained by the SM.

\subsection{Solar neutrinos production}
Based on our understanding of stellar evolution we know that nuclear
reactions fuel the Sun. Nuclear processes involve only the first
generation of fermions and thus only electron neutrinos are generated
in the Sun. Moreover, since the Sun is burn by fusion reactions it
produces only neutrinos and not anti-neutrinos. The neutrino energies
are relatively small, $E_\nu \lsim 10\;{\rm MeV}$.  Therefore, only
disappearance experiments can be done; the neutrinos energies are too
small to produce muons or taus in any target at rest.

There are many processes that produce solar neutrinos. The main
reaction chain, which produces about $98.5\%$ of the solar energy (and
most of the solar neutrinos) is the $pp$ cycle
\beq
4p \to \, ^4{\rm He} + 2 e^+ + 2 \nu_e + 2 \gamma\,.
\eeq
The neutrinos emitted in this cycle have continuum spectrum which
goes up to $E_\nu < 0.42\;{\rm MeV}$.  Assuming that the Sun is in
equilibrium, one can use its current luminosity to calculate the
neutrino flux from the $pp$ reaction chain.

The other nuclear reactions in the Sun are not very important for its
luminosity, but they are important for neutrino physics.  The
difference originates from the fact that photons are thermalized
almost immediately after they are produced. Neutrinos, on the other
hand, do not interact in the Sun, and leave the Sun with their
original energies.  Several sub-dominant reactions produce neutrinos
with energies above the threshold of the $pp$ chain. Since it is
easier to detect high energy neutrinos, in many cases the experimental
signals consist only of neutrinos that were produced by sub-dominant
reactions.

\begin{figure}[t]
\mycenterline{~~~~~\includegraphics[height=13cm,angle=-90]{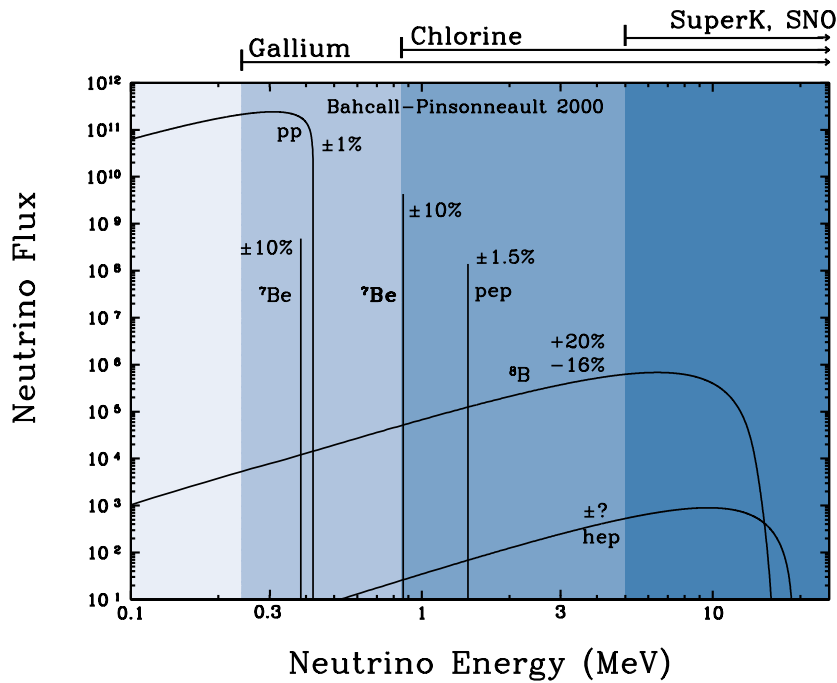}}
\caption[b]{The Bahcall and Pinsonneault 2000  solar model prediction
of the neutrino flux as function of the neutrino energy. This
figure was taken from \cite{bahcall-web}, where many more details on
solar neutrinos can be found.\label{fig-spec}}
\end{figure}

While the spectra of the different reactions are known, the
sub-dominant reaction chains fluxes are not known in a model
independent way.  Solar models, namely, theoretical arguments, are
used to obtain the total neutrino spectrum. As an example, the Bahcall
and Pinsonneault 2000 (BP00) solar model prediction for the neutrino
flux is given in fig.~\ref{fig-spec}.  Since solar models involve
theoretical inputs, one would like to perform experiments where the
interpretation of the data can be done with minimum dependence on a
specific solar model.

\subsection{Solar neutrinos propagation}

Solar neutrinos are produced in the vicinity of the solar core. Then,
they freely propagate out of the Sun and some of them travel to our
detectors.  During the night, neutrinos reach the detector after they
have passed through the Earth. It is convenient to separate the solar
neutrino propagation into three regions: inside the Sun, between the
Sun and the Earth, and inside the Earth. We discuss each of these
below.

As already mentioned before, matter effects in the Sun can be
important for flavor conversion. To calculate these effects we need to
know the matter profile in the Sun and the rates of the different
reactions in the different parts of the Sun.  While these parameters
are roughly known, there are uncertainties in their values. Therefore,
we have to count on solar models to provide the needed inputs.

The distance to the Sun is known to high precision. It varies by about
$7\%$ during the year. If vacuum oscillation is relevant to solar
neutrinos then this distance variation can be used in order to get
better control on the neutrino parameters.

During the night neutrinos pass through the Earth on their way to the
detector, while during the day they do not.  Therefore, the Earth matter
effect, which shows up as a day-night difference of the detected
neutrino flux, provides further sensitivity to the neutrino
parameters. Moreover, depending on the location of the detector and
the time of the year the neutrino may pass through both the Earth
mantle and its core or only through the mantle. This is important since the
matter density in these two regions is different, and therefore can
affect the oscillations \cite{Akhmedov}.

\subsection{Solar neutrinos detections}

There are three reactions that are used to detect solar neutrinos.
The Charged Current (CC) interaction 
\beq
\nu_e + n \to p + e^-\,,
\eeq
where $n$ is a neutron, can be used to detect only electron neutrinos.
All the neutrinos can undergo Elastic Scattering (ES)
\beq
\nu_\ell + e^- \to \nu_\ell + e^-\,.
\eeq
This reaction is mediated by neutral current $Z$ exchange for all
neutrino flavors. For electron neutrinos there is also contribution
from a $W$ exchange amplitude.  Due to this difference, the elastic
scattering cross section for electron neutrinos is about six times
larger than for muon and tau neutrinos.  Finally, the cross section
for the Neutral Current (NC) interaction
\beq
\nu_\ell + n \to \nu_\ell + n\,,
\eeq
is the same for all neutrino flavors (here $n$ is a neutron).

When presenting results of neutrino oscillation experiment it is
convenient to define
\beq
R=N_{obs}/N_{MC}\,,
\eeq 
such that $N_{obs}$ is the number of detected neutrinos and $N_{MC}$
is the number of events predicted from a solar model based Monte Carlo
(MC) simulation assuming no flavor oscillation.  For detectors that
use neutral current reactions $R$ should be one regardless if there
are active neutrino flavor oscillations or not. For charged current
and elastic scattering based detectors, however, $R=1$ is expected
only if there are no neutrino flavor oscillations, and $R<1$ is
expected if there are oscillations.

There are several solar neutrino experiments that use the above
mentioned basic reactions. They have different experimental setups
that us various reactions with different sensitivities and
thresholds.  Next we describe these experiments.

\subsection{Chlorine: Homestake}
The first detection of solar neutrinos was announced 
35 years ago. The neutrinos were detected 
in the Homestake mine in South Dakota \cite{davis}. 
The experiment uses the
\beq
\nu_e + {}^{37}{\rm Cl} \to  {}^{37}\!{\rm Ar} + e^-\,,
\eeq
charged current interaction in order to capture the electron neutrinos.

We note the following points: 
\bim
\item
The signal is extracted off-line. Namely, only integrated rates are
measured. Therefore, no day-night and almost no spectral information
are available. (The only information about the neutrino spectrum is
that the energy of the neutrino is above the threshold.)
\item
The energy threshold is $0.814$ MeV. Therefore, the experiment is
sensitive mainly to the $^7{\rm Be}$ and $^8{\rm B}$ neutrinos. $pp$
neutrinos cannot be detected in this experiment.
\item
After many years of data collection the published result is \cite{homestake1}
\beq \label{cl-data}
R\approx 0.3\,.
\eeq
Of course, in order to perform a fit to the neutrino parameters the
above crude approximation should not be used. Yet, quoting the order
of magnitude, as we did in eq. (\ref{cl-data}) is sufficient to
understand the main ingredients of the solar neutrino problem.
\eim
The fact that the Homestake experiment found $R<1$ was the first
indication for solar electron neutrino disappearance. This result gave
the motivation to build different types of detectors in order to
further study the solar neutrinos.

\subsection{Gallium: SAGE, GALLEX and GNO}
The Homestake experiment had relatively high threshold.  The SAGE, 
GALLEX \cite{GALLEX} and GNO \cite{GNO} Gallium
experiments were built in order to detect neutrinos with much smaller
energies. In particular, their threshold is low enough to detect $pp$
neutrinos.  
In the Gallium experiments the neutrinos are detected using the
\beq
\nu_e + {}^{71}{\rm Ga} \to {}^{71}{\rm Ge} + e^-
\eeq
charged current reaction.

We note the following points:
\bim
\item
Like the Chlorine experiment, the signal is extracted off-line. Thus,
also here no day-night and spectral information are available.
\item
The energy threshold is $0.233$ MeV. Thus, Gallium detectors are
sensitive to $pp$ neutrinos. The $^7{\rm Be}$ and $^8{\rm B}$
neutrinos also contribute significantly to the neutrino signal.
Therefore, the total predicted flux is not solar model independent.
\item
Combining the results of the three experiments, one obtains
\beq
R\approx 0.6.
\eeq
\eim
We learn that like in the Chlorine experiment, also here, the number
of observed electron neutrinos is less than predicted.  Moreover, the
measured $R$ is different for the two type of experiments. Since these
two types are sensitive to different energy ranges, this fact
indicates that the electron neutrino flux reduction must be energy
dependent.

\subsection{Water Cerenkov: Kamiokande and SuperKamiokande}

Both the Chlorine and Gallium experiments extract their signal
off-line. The Kamiokande and SuperKamiokande \cite{SK} water Cerenkov
detectors can extract their signal on-line. Therefore, in these
experiments day-night and spectral information can be extracted. The
price to pay is that the energy threshold is much higher compared to
the thresholds of the Chlorine and Gallium experiments.

In the water Cerenkov detectors the neutrinos are detected by elastic
scattering of the neutrinos off the electrons in the water
\beq
\nu_\alpha + e^- \to  \nu_\alpha+ e^-\,.
\eeq

We note the following points:
\bim
\item
Since the detection is done on-line, the time of each event is known,
and also some information about the direction and energy of the
neutrino in each event can be extracted. These features enable the
experiments to clearly show that the detected 
neutrinos are coming from the Sun.
\item
The water Cerenkov detectors threshold is about $6$ MeV.  Thus, the
Kamiokande and SuperKamiokande experiments are sensitive mainly to $^8 {\rm B}$
neutrinos.
\item
Since elastic scattering is used for the detection, all neutrino
flavors contribute to the signal.  This fact is important if there are
active flavor oscillations.  Then, the converted neutrinos also
contribute to the signal. This is in contrast to detectors that use
charged current interactions where only electron neutrinos generate
the signal.
\item
The integrated suppression is found to be 
\beq
R\approx 0.5.
\eeq
Spectrum and day-night information were also reported.
\eim
We learn again that there is an energy dependent suppression of the
electron neutrino flux.

\subsection{Heavy Water: SNO}
We already mentioned that neutrino flavor conversion is the most
plausible explanation to the observed energy dependent electron
neutrino flux suppression.  Thus, it is interesting to measure solar
neutrinos using neutral current reactions since in that case all
neutrino flavors contribute with the same strength. Namely, neutral
current based experiments measure the solar neutrino flux independent
of whether there are active flavor oscillations or not.

The SNO detector \cite{SNO} was built for this purpose. Its apparatus
contains an inner tank filled with heavy water and an outer tank
filled with regular water. Neutrino can be detected by all the three
basic reaction types
\beqa
{\rm CC}(E_{th} = 2.23\,{\rm MeV}):&&\quad \nu_e+^2{\rm \!H}\to p+p+e^- \,,
\nonumber \\
{\rm ES}(E_{th} \sim 6\,{\rm MeV}):&& \quad 
 \nu_\alpha + e^- \to \nu_\alpha+e^- \,, \nonumber \\
{\rm NC}(E_{th} \sim 6\,{\rm MeV}):&& \quad 
 \nu_\alpha + ^2{\rm \!H} \to n+p+\nu_\alpha \,,
\eeqa
where $E_{th}$ is the threshold energy. Note that the presence of
Deuterium ($^2$H) in the heavy water is crucial for the charged
current and neutral current measurements.

\begin{figure}[t]
\mycenterline{\includegraphics[height=8cm]{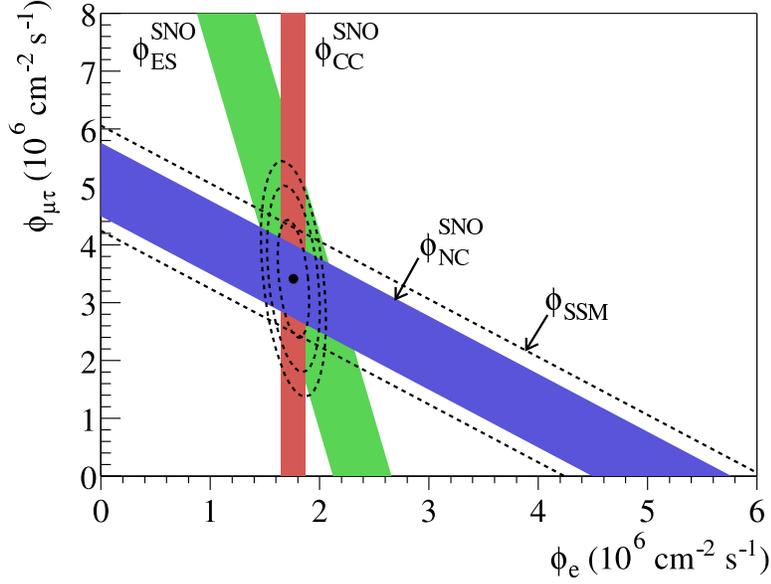}}
\caption[c]{The SNO data is plotted as three independent 
measurements of the muon or tau neutrino flux ($\phi_{\mu\tau}$) as a
function of the electron neutrino flux ($\phi_e$).  The three bands
represent the neutral current, charged current and elastic scattering
based measurements. The dotted line is the BP00 solar model prediction for
the neutrino flux. It is plotted only in conjunction with the neutral
current measurement since only this measurement is insensitive to
electron to muon or tau flavor conversion. (This plot is taken from
\cite{sno-result}.)
\label{fig-sno3}}
\end{figure}

Note the following points:
\bim
\item
The signal is extracted on-line. Thus day-night and spectrum
information is available.
\item
We define the ratios 
\beq
R_E \equiv {R_{\rm CC} \over R_{\rm ES}}, \qquad
R_N \equiv {R_{\rm CC} \over R_{\rm NC}},
\eeq
where $R_{\rm CC}$ ($R_{\rm ES}$, $R_{\rm NC}$) is the ratio between
the observed number of charged current (elastic scattering, neutral
current) events and the predicted number assuming no oscillation.
Extracting $R_E$ and $R_N$ has several advantages. First, many
systematic uncertainties cancel in their measurements. Second, their
predicted values are almost solar model independent and depend only on
the neutrino parameters.  In particular, $R_E \ne 1$ or $R_N \ne 1$
can occur only if electron neutrinos are converted to muon or tau
neutrinos.
\item
SNO found $R_E \ne 1$ and $R_N \ne 1$.  This result confirms the
existence of $\nu_\mu$ or $\nu_\tau$ component in the solar neutrino
flux at $5.3\sigma$. In addition, the BP00 solar model prediction is
confirmed by the neutral current measurement. The SNO result
\cite{sno-result} is illustrated in fig.~\ref{fig-sno3}.
\eim

\subsection{Solar neutrinos: fits}

The general picture emerging by combining all the solar neutrino data
is as follows.  All charged current and elastic scattering based
measurements found less events than predicted for massless
neutrinos. This suppression is energy dependent.  The BP00 solar model
prediction for the neutral current based measurement, which holds also
if there are active neutrino flavor conversions, was confirmed by SNO.

The first and most robust conclusion is that the SM, which predicts
massless neutrinos, fails to accommodate the data.  While there are
several exotic explanations that may be able to explain the data
\cite{exotic}, the simplest and most plausible one is that of active
neutrino flavor oscillation
\beq
\nu_e \to \nu_x, \qquad x=\mu,\tau\,.
\eeq
In that case, experiments that use reactions that are flavor
dependent, namely charged current interactions and elastic scattering,
are expected to detect fewer events than anticipated.  On the other
hand, experiments that use neutral current interactions, which are
flavor blind, should not observe such reduction. Indeed, this is what
the data shows.

\begin{figure}[t]
\mycenterline{\includegraphics[height=12cm]{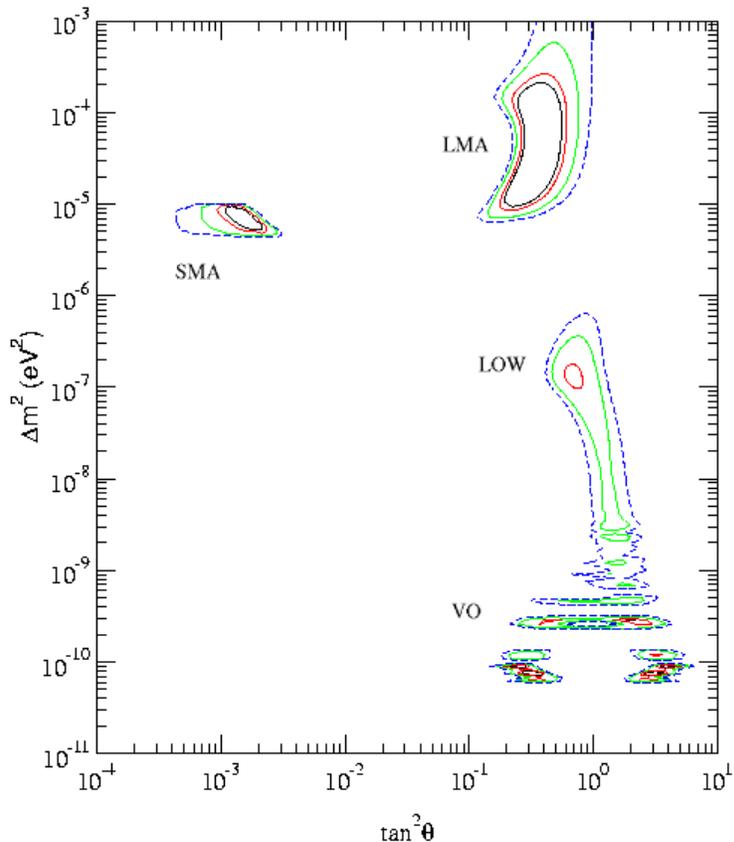}}
\caption[d]{An example of a neutrino parameters fit to some 
solar neutrino data.  The names of the regions where good fit is found
are indicated in the figure. Note that by now this fit is outdated, and
it is shown here for illustration only. (This plot is based on
a plot taken from \cite{Bandyopadhyay}.)
\label{SN-fit}}
\end{figure}

Assuming that the neutrinos are massive, the data can be used to
determine the neutrino masses and mixing angles. While three flavor
oscillation fits are available \cite{fits}, it is a good approximation
to fit the data assuming only two neutrino mixing.  An example of such
fit is shown in fig. \ref{SN-fit}.  (Recent fits can be found in
\cite{fits}. Note, however, that whenever new data is published, new
fits are performed and published usually within days.)  The regions
where a good fit is found are indicated in the figure. The best fit is
obtained in the Large Mixing Angle (LMA) region. The recent SNO data
actually exclude the Small Mixing Angle (SMA) region. The LOW and the
Vacuum Oscillation (VO) regions, are still allowed by all solar
data. They are excluded, however, by the recent KamLAND result, see
below.

To conclude, there are strong indications for active $\nu_e \to \nu_x$
oscillation. This conclusion is robust since it has practically no
dependence on the solar model. The best fit to the neutrino parameters
is in the LMA region with
\beq \label{SNfit}
\Delta m^2 = 6 \times 10^{-5}\,{\rm eV}^2, \qquad
\tan^2 \theta =0.4.
\eeq
Future data and new experiments will teach us more about solar
neutrinos. Besides having more statistics, future experiments will
measure observables where we have only very little information at
present.  In particular, we expect to have more information about the
day-night effect and the low energy neutrino spectrum.

\section{Atmospheric neutrinos}
In the reaction chain initiated by cosmic ray collisions off nuclei in
the upper atmosphere neutrinos are produced. These neutrinos are
searched for by atmospheric neutrinos experiments. There are
indications for muon neutrinos disappearance.  These indications are
called the atmospheric neutrino problem or the atmospheric neutrino
anomaly. The likely explanation of this phenomenon is active neutrino
oscillation.

\subsection{Atmospheric neutrino production}
Cosmic ray collisions off nuclei in the upper atmosphere produce
hadrons, mainly pions. In their decay chain 
\beq \label{sff}
\pi^+ \to \mu^+ \nu_\mu, \qquad
\mu^+ \to e^+ \nu_e \,\overline \nu_\mu, 
\eeq
(and the charge conjugated decay chain) neutrinos are produced. Note
that almost exclusively electron and muon neutrinos are
produced. Based on (\ref{sff}) we expect
\beq
N(\nu_\mu)/N(\nu_e)\approx 2,
\eeq
where $N(\nu_\ell)$ is the number of neutrinos of flavor $\ell$. There
are corrections to this prediction. Other mesons, mainly kaons, are
also produced by cosmic rays.  Their decay chains produce different
neutrino flavor ratio. Another effect is that at high energies the
muon lifetime in the lab frame is much longer than in its rest frame.
In fact, it is long enough such that not all the muons decay before
they arrive to the detector. All in all, the neutrino production rates
and spectra are known to an accuracy of about $20\%$. The ratio
$N(\nu_\mu)/N(\nu_e)$ is known better, to an accuracy of about $5\%$.

The atmospheric neutrino energies are relatively large, $E_\nu \gsim
100$ MeV. Therefore, both disappearance and appearance experiments are
possible. In particular, since the initial tau neutrino flux is tiny,
searching for tau appearance is an attractive option.

\subsection{Atmospheric neutrino propagation}
Since neutrinos are produced isotropically in the upper atmosphere,
they reach the detector from all directions. The neutrino zenith angle
in the detector is correlated with the distance it traveled.
Neutrinos with small zenith angle, namely those coming from above,
travel short distances (of order $10^2\;$km). Neutrinos with large
zenith angle, namely those coming from below, are neutrinos that cross
the Earth and travel much longer distances (of order $10^4\;$km).
Since the production is isotropic, the neutrino flux in the detector
is expected to be isotropic in the absence of oscillation. (There are
small known corrections due to geo-magnetic effects.) 

Matter effects can be important for neutrinos that cross the Earth.
Oscillations that involve electron neutrinos
or oscillations between active and sterile neutrinos are modified in
the presence of matter.  In contrast, the matter has almost no effect
effect on $\nu_\mu \to \nu_\tau$ oscillation.


\subsection{Atmospheric neutrino detection}

There are two types of atmospheric neutrino detectors. The water
Cerenkov detectors, IMB, Kamiokande and SuperKamiokande \cite{SK}, use
large tanks of water as targets.  The water is surrounded by
photomultipliers which detect the Cerenkov light emitted by charged
leptons that are produced in charged current interactions between the
neutrinos and the water.  Compared to electrons, muons are likely to
create sharper Cerenkov rings.  Thus, the shape of the ring helps in
determining the flavor of the incoming neutrino.  Iron calorimeter
detectors are based on a set of layers of iron which act as a target,
and some tracing elements that are used to reconstruct muon tracks or
showers produced by electrons.

Both water Cerenkov and Iron calorimeter detectors identify the
neutrino flavor via its charged current interactions.  They have some
ability to look for neutral current interactions.  Information on the
neutrino energy and direction can also be extracted by both methods.

\subsection{Atmospheric neutrino data}

\begin{figure}[t]
\mycenterline{\includegraphics[height=5cm]{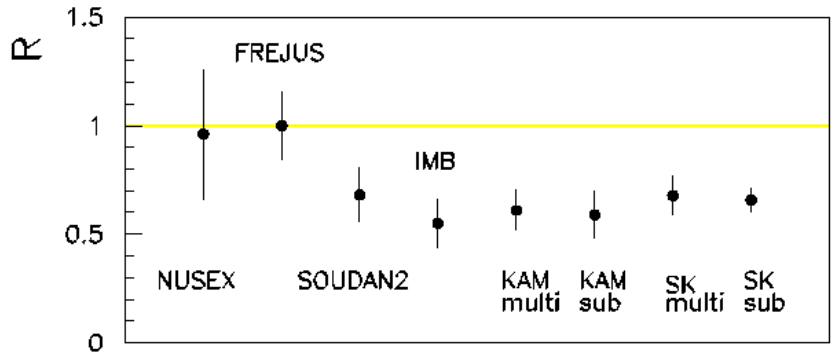}}
\caption[e]{Summary of the experimental measurements of $R$ defined in eq.
(\ref{defR}).  (This plot is taken from \cite{cy}.)
\label{fig-cont}}
\end{figure}

All atmospheric neutrino experiments measure the double ratio
\beq \label{defR}
R = {N^\mu_{obs}/N^e_{obs} \over N_{MC}^\mu/N_{MC}^e},
\eeq
where $N^\ell_{MC}$ is the expected number of flavor $\ell$ type
events and $N^\ell_{obs}$ are the observed ones. The advantage of
using this double ratio is that many systematic and theoretical errors
cancel in this ratio.  Almost all of the experiments found $R<1$; see
fig.~\ref{fig-cont}.

\begin{figure}[t]
\ifnum\vertou=0
\mycenterline{\includegraphics[height=6.6cm]{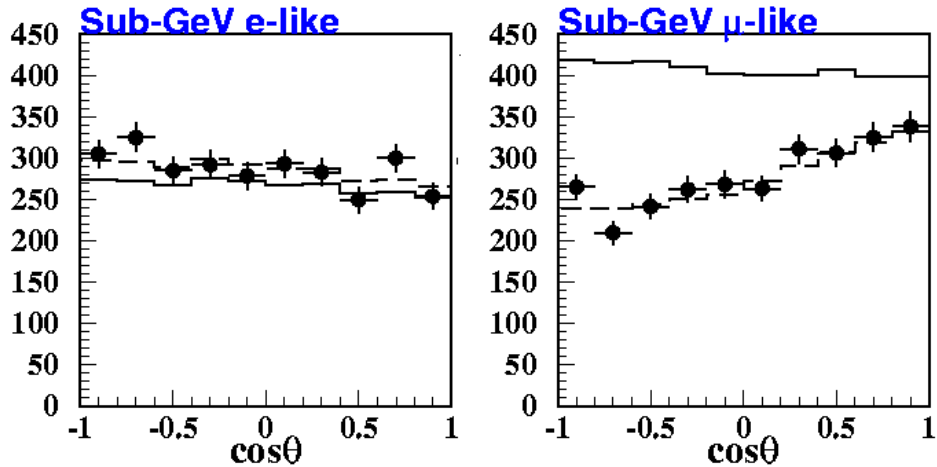}}
\else
\mycenterline{~~\includegraphics[height=5.9cm]{ANs.ps}}
\fi
\caption[f]{The low energy electron and muon neutrino fluxes 
compared to the MC predictions as observed by SuperKamiokande. 
The solid lines are the expected fluxes assuming no
oscillations. The dashed lines are the best fits to neutrino
oscillation. The solid circles are the data points. (This plot is
taken from \cite{SKplot}.) \label{fig-ANs}}
\end{figure}

The observation of $R<1$ can be explained by muon neutrino
disappearance, electron neutrino appearance, or both.  The
SuperKamiokande experiment also measured the zenith angle dependence
of the neutrino flux. The data indicate that the preferred explanation
is that of muon neutrino disappearance. For example,
fig.~\ref{fig-ANs} shows that the low energy $\nu_e$ flux agrees while
the $\nu_\mu$ flux disagrees with the MC prediction.

\subsection{The atmospheric neutrino problem}
The zenith angle dependence of the muon neutrino flux as well as $R<1$
cannot be explained with SM massless neutrinos. This is called the
atmospheric neutrino anomaly.  In order to explain it disappearance of
muon neutrinos is needed. While there are several exotic explanations
\cite{exotic}, the most attractive explanation is active neutrino
flavor oscillation.  In particular, the best fit is achieved for
$\nu_\mu \to \nu_\tau$ oscillation with
\beq \label{ANfit}
\Delta m^2=2.6 \times 10^{-3} \;{\rm eV}^2, \qquad
\sin^2 2 \theta =  0.97.
\eeq

The hypothesis of neutrino oscillation can be further tested. First,
terrestrial long baseline experiments probe the same region of
parameter space as the atmospheric neutrino experiments. Indeed, as
discussed in more detail below, the recent K2K long baseline
experiment results agree with the atmospheric neutrino data. Second,
search for tau appearance in the atmospheric neutrino data is also
possible. The SuperKamiokande experiment found such indications
\cite{SK-tau}, but they are not yet convincing.

\section{Terrestrial neutrinos}
One disadvantage of solar and atmospheric neutrino measurements is
that since the neutrinos are not produced on Earth, there is no
control over the neutrino source. Neutrinos from accelerators and
nuclear reactors have the advantage that their source is known. That
is, we know to high accuracy the initial neutrino flux, the energy
spectrum and the flavor composition of the neutrino source. Moreover,
in some cases these parameters can be tuned in order to maximize the
experimental sensitivity.

The main advantage of solar and atmospheric neutrinos is that they are
sensitive to very small $\Delta m^2$. That is, their sensitivity is
much better compared to most terrestrial neutrino
experiments. Therefore, it is not surprising that most terrestrial
neutrino experiments did not find disappearance or appearance
signals. Their data was used to set bounds on the neutrino parameters.
There are, however, some exceptions. The sensitivity of the K2K
\cite{K2K} and KamLAND \cite{Kamland} experiments are similar to those
of, respectively, the atmospheric and solar neutrino
experiments. Recently, positive disappearance signals were found in
these two terrestrial experiments.  The situation is different with
regard to the LSND \cite{LSND} experiment. In that case there is a
claim for an appearance signal. This signal can only be explained with
neutrino masses that are much larger compared to the masses deduced
from the solar and atmospheric neutrino data. Next we describe these
three experiments and their results.

%
\subsection{K2K}
In order for accelerator neutrinos to be sensitive to the same range
of parameter space as the atmospheric neutrino experiments, long
baseline, of order $10^3$ km, is needed.  In such long baseline
experiments it is possible to search for $\nu_\mu$ disappearance or
$\nu_\tau$ appearance. In these experiments a neutrino beam from an
accelerator is aimed at a detector located far away.  The first
operating long baseline experiment is the K2K experiment. It uses an
almost pure $\nu_{\mu}$ beam that is generated at KEK and is detected
at SuperKamiokande, which is about 250 km away.

Recently, the K2K experiment announced their first result
\cite{K2Kdata}. 
They observed 56 muon neutrino events with an expectation of about
80. The $\nu_\mu$ disappearance can be explained by $\nu_\mu \to \nu_x$
flavor oscillation. The best fit is found for the following values
\beq
\Delta m^2 = 2.8 \times 10^{-3}\; {\rm eV}^2, \qquad
\sin^2 2 \theta = 1.0.
\eeq
These values of the neutrino parameters are very close to those
indicated by the atmospheric neutrino data, see eq. (\ref{ANfit}). The
statistical significance of the K2K oscillation signal is smaller than
the atmospheric neutrino ones. The importance of the K2K result lies
in the fact that it provides an independent test of the atmospheric
neutrino parameter space.

%
\subsection{KamLAND}
Reactor neutrinos have much smaller energies compared to neutrinos
produced at accelerator. Thus, long baseline reactor experiments are
sensitive to smaller values of $\Delta m^2$.  Another important
difference is that reactors generate only electron anti-neutrinos,
while accelerators produce mainly muon neutrinos. Thus, reactor
experiments can be used to provide independent probes of the solar
neutrino parameter space.
 
The KamLAND experiment is placed in the Kamioka mine in Japan.  This
site is located at an average distance of about 180 km from several
large Japanese nuclear power stations.  The initial fluxes and spectra
of the $\overline\nu_e$ emitted in each reactor are known to a good
accuracy, since they are related to the power that is generated in
each reactor.  Thus, measurement of the total flux and energy spectrum
of the $\overline\nu_e$ at KamLAND can be used to search for
disappearance.  This is particularly interesting since the KamLAND setup
is such that it is sensitive to the region of the parameter space 
indicated by the LMA solution to the solar neutrino anomaly.

Recently, the KamLAND experiment announced their first result
\cite{kamland-data}. 
They found 
\beq
{N_{obs} \over N_{MC}} = 0.611 \pm 0.085 \pm 0.041
\eeq
where $N_{obs}$ is the number of observed events and $N_{MC}$ is the
expected  number. This result
can be explained by $\overline\nu_e \to \overline\nu_x$ 
neutrino flavor oscillation with the best fit at 
\beq
\Delta m^2 = 6.9 \times 10^{-5}\; {\rm eV}^2, \qquad
\sin^2 2 \theta = 1.0.
\eeq
These values are very close to the ones indicated by the solar
neutrino data, see eq. (\ref{SNfit}).

The importance of this result is twofold. First, it provides a
completely solar model independent test for the neutrino oscillation
solutions of the solar neutrino problem. Moreover, it discriminates
between the different possible solutions, pointing at the LMA as the
favorable one.

%
\subsection{LSND}

The set up of the LSND experiment \cite{LSND} is as follows.
Neutrinos are produced by sending protons on a fixed target that
generates pions. Almost all the negatively charged pions are absorbed
in the target before they decay. The neutrinos are produced via
\beq
\pi^+ \to \mu^+ \nu_\mu, \qquad
\mu^+ \to e^+ \nu_e \overline\nu_\mu.
\eeq
Then, the neutrinos travel for about 
30 meter to the detector.
Since the  beam contains $ \nu_\mu$,  $\nu_e$ and $\overline\nu_\mu$, with
negligible fraction of $\overline\nu_e$, it best to search for
$\overline\nu_\mu \to \overline\nu_e$ appearance.

The LSND collaboration announced a $3.3\sigma$ indication for
$\overline \nu_e$
appearance
\beq
P(\overline\nu_\mu \to \overline\nu_e) = 
(2.64 \pm 0.67 \pm 0.45) \times 10^{-3}.
\eeq
This result can 
be explained with  $\overline\nu_\mu \to \overline\nu_e$ oscillation
where the best fit is achieved for
\beq \label{LSND-dat}
\Delta m^2 =1.2~{\rm eV}^2, \qquad \sin^2 2\theta = 0.003.
\eeq
Note that the value of $\Delta m^2$ in (\ref{LSND-dat})
is much larger than the values
indicated by the solar and atmospheric neutrino data.

The LSND indications for neutrino masses are not as strong as the
solar and atmospheric neutrino ones. First, the statistical
significance of the LSND signal is rather low (traditionally, only a
$5\sigma$ effect is called a discovery). Second, the LSND signal still
needs an independent confirmation.  The Karmen experiment
\cite{Karmen}, which has a setting similar to LSND, but with
a shorter baseline, $L\approx 17$ meter, did not find an appearance
signal. Soon, the MiniBooNE experiment \cite{miniboone} will be able
to clarify the situation.

\section{Theoretical implications}
As we have seen, there are solid experimental indications for
neutrino masses. The most important implication of these results is also
the most simple one: the SM is not the complete picture of
Nature. Namely, there are experimental indications that the SM is
only an effective low energy description of Nature.

There are basically two ways to extend the SM. One way is to assume
that there is new physics only at a very high scale, much above the weak
scale. Then, neutrino masses can be accounted for by the seesaw
mechanism.  Alternatively, one can think about weak scale new physics.
Neutrino masses that are generated by such new physics are, in general,
too large and some mechanism is required in order to explain the
smallness of the neutrino masses.
In the following we concentrate on the former option.

\subsection{Two neutrino  mixing}

We start by considering only one type of results at a time. Namely, we
draw conclusions from the solar and KamLAND data or from the
atmospheric and K2K data.  In that case there is one theoretical
challenge imposed by the data: how to generate neutrino masses
at the right scale?

The atmospheric neutrinos and K2K results indicate $\nu_\mu \to
\nu_\tau$ oscillation with
\beq
\Delta m^2_{\rm AN} \sim {\rm few}\times 10^{-3}\,{\rm eV}^2.
\eeq
The solar neutrinos and KamLAND results indicate $\nu_e \to
\nu_{\mu,\tau}$ oscillation with
\beq
\Delta m^2_{\rm SN}  \sim 10^{-4} \,{\rm eV}^2.
\eeq
We use (\ref{genmass}) in order to
obtain the scale of the new physics that is responsible for neutrino
masses
\beq
m_\nu = {m_D^2 \over M} \quad \gorer \quad 
M = {m_D^2 \over m_\nu}.
\eeq
In order to find the high energy scale, $M$, we need to know $m_\nu$
and $m_D$.  The most likely situation is that the neutrinos are not
degenerate \beq {\Delta m_{ij}^2 \over m_i^2+m_j^2} \sim 1.
\eeq
Then, we use $m_\nu \sim \sqrt{\Delta m^2}$.  In general, the
Dirac masses are of the order of the weak scale times 
dimensionless couplings. The problem is that we do not know the value
of these couplings. In the following we make two plausible choices for
their values. Motivated by $SO(10)$ GUT theories, we assume that these
couplings are of order of the up type quark Yukawa couplings.  In
particular, for the heaviest neutrino, the coupling is of the order of the
top Yukawa coupling. Another plausible choice is to use the charged
lepton Yukawa couplings as a guide to the neutrino ones.

Considering atmospheric neutrino and K2K data we use $m_\nu\sim {\rm
few}\times 10^{-2}\,{\rm eV}$ and get
\beq \label{eq-an-scale}
\begin{array}{ll}
m_D \sim m_t &\quad \gorer \quad  M \sim 10^{16} {\rm ~GeV}, \\
m_D \sim m_\tau & \quad \gorer \quad    M \sim 10^{11} {\rm ~GeV}. 
\end{array}
\eeq
If instead we use  $m_\nu \sim 10^{-2}\,{\rm eV}$ as indicated by
solar neutrinos  and KamLAND data, we obtain
\beq\label{eq-sn-scale}
\begin{array}{ll}
m_D \sim m_t & \quad \gorer \quad  M  
\sim {\rm few} \times 10^{16} {\rm ~GeV}, \\
m_D \sim m_\tau &  \quad \gorer \quad   M \sim 10^{12} {\rm ~GeV} .
\end{array}
\eeq

We emphasize the following points:
\bim
\item
Both eqs. (\ref{eq-an-scale}) and (\ref{eq-sn-scale}) imply that there
exists a new scale between the weak scale and $\Mpl \sim 10^{19} {\rm
~GeV}$. 
\item
The values obtained assuming $m_D \sim m_t$ in both
eqs. (\ref{eq-an-scale}) and (\ref{eq-sn-scale}) are very close to the
GUT scale, $\Mgut \sim 3\times 10^{16} {\rm ~GeV}$.  Therefore, we can
say that neutrino masses are indications in favor of unification.
\item
In several new physics models, a new scale, usually called the
intermediate scale, $\Mint \sim \sqrt{m_W \Mpl}\sim 10^{11} {\rm
~GeV}$, is introduced. For example, supersymmetry breaking has to
occur at this scale if it is mediated via Planck scale physics to the
observable sector.  The values obtained assuming $m_D \sim m_\tau$ in
both eqs. (\ref{eq-an-scale}) and (\ref{eq-sn-scale}) are very close
to $\Mint$. This could be an indication that neutrino masses are also
generated by such models. 
\eim

%
\subsection{Three neutrino mixing}

When we combine the solar, KamLAND, atmospheric
and K2K data, we have to consider the full three flavor
mixing. There are three mass differences that control the
oscillations. They are subject to one constraint
\beq \label{cons}
\Delta m^2_{12}+ \Delta m^2_{23} + \Delta m^2_{31}=0\,.
\eeq
The data indicates that the mass-squared differences are hierarchical,
$|\Delta m^2_{12}| \ll |\Delta m^2_{23}|$. This implies that $|\Delta
m^2_{23}| \approx |\Delta m^2_{31}|$, and therefore, that there are
only two different oscillation periods.

There are three mixing angles.  Solar neutrino oscillations depend
mainly on $\theta_{12}$ and atmospheric neutrino oscillations depend
mainly on $\theta_{23}$. The third angle, $\theta_{13}$, is known to
be small, mainly from terrestrial experiments (see, for example,
\cite{cy} for details).  Thus, the oscillation phenomena are described
by two mass-squared differences and three mixing angles
\beqa \label{retionpp}
&&
\Delta m^2_{12} \sim 10^{-4}\,{\rm eV}^2, \qquad
\Delta m^2_{23} \sim {\rm few}  \times 10^{-3}\,{\rm eV}^2, \\
&&
\theta_{12} \sim 1, \qquad\qquad
\theta_{23} \sim 1, \qquad\qquad
\theta_{13} \lsim 0.2. \nonumber 
\eeqa
As before, we assume that the neutrinos are not degenerate. Then, we
learn from (\ref{retionpp}) that the neutrino masses are somewhat
hierarchical with two large and one small mixing angles.

The theoretical challenge imposed by considering three neutrino flavor
mixing is to explain the flavor structure of (\ref{retionpp}).  There
are basically two approaches to do so.  One is called neutrino anarchy
\cite{anarchy}. It assumes that there are no parametrically small
numbers. The two apparently small numbers in the neutrino sector
\beq
{m_2 \over m_3}  \sim {1 \over 5}, \qquad
\theta_{13} \sim {1 \over 5},
\eeq
are assumed to be accidentally small. That is, 
numbers of order five are considered natural. Therefore, this mechanism
predicts that $\theta_{13}$ is close to its current upper
bound. Consequently, it also predicts that CP violating observables,
which are proportional to the product of all the mixing angles and the
neutrino mass-squared differences, should also be close to their
current upper bounds.

The other option to explain (\ref{retionpp}) is to assume that there
is an underlying broken flavor symmetry that controls the neutrino
sector parameters. Such a symmetry is often assumed in order to
explain the quark sector flavor parameters. Within this framework it
is assumed that there is one small parameter, $\epsilon$,
such that all the small parameters of the theory are functions of
it. In particular,
\beq
{m_2 \over m_3} \sim \epsilon^{n} \ll 1 , \qquad
\theta_{13} \sim \epsilon^{m} \ll 1,
\eeq
where $n$ and $m$ are some positive integers.  In that case we often
refer to observables like $m_2 / m_3$ and $\theta_{13}$ as
parametrically small numbers.

While flavor symmetries can explain the observed hierarchies in the
quark sector, it is not straightforward to extend the treatment to the
neutrinos. The main reason is that in general large mixing angles come
without mass hierarchies and vice versa. Explicitly, we consider the
two generation case. The neutrino mass matrix is
\beq
m =  \pmatrix{a & b \cr b & c},
\eeq
and we distinguish the following three cases
\bim
\item 
 $a \gg b,c ~\gorer~ \sin^2\theta \ll 1, \quad {m_1 / m_2} \ll 1 $.
\item
$a,b,c \sim 1,~~\det(m) \sim 1 ~\gorer ~\sin^2\theta \sim 1,~~
{m_1 / m_2} \sim 1 $.
\item
$a,b,c \sim 1,~~\det(m) \ll 1 ~\gorer ~\sin^2\theta \sim 1,~~ {m_1 /
m_2} \ll 1$. 
\eim  
We learn that in order to have large mixing with mass hierarchy the
determinant of the mass matrix must be much smaller than the typical
value of its entries (to the appropriate power). This is not a common
situation in flavor models. In particular, this is not the case in the
quark sector.

%
\subsection{Four neutrino mixing}

While the LSND data cannot be considered as a solid indication for
neutrino masses, one may like to try to accommodate it as well.  In
that case the situation become much more complicated.  The reason is
that three different mass-squared differences are needed
\beq
\Delta m^2_{\rm SN}\sim 10^{-4} \,{\rm eV}^2, \quad
\Delta m^2_{\rm AN} \sim {\rm few} \times 10^{-3} \,{\rm eV}^2, \quad 
\Delta m^2_{\rm LSND} \sim 1 \,{\rm eV}^2. 
\eeq
With three neutrinos, however, the constraint (\ref{cons}) implies
that there are at most two different scales.  Therefore, at least four
neutrinos are needed in order to accommodate all of these results.
Experimentally we know that there are only three active neutrinos, and
thus we need at least one additional light sterile neutrino field.

There are two theoretical challenges: First, a sterile neutrino, by
definition, is a fermion that is a singlet under the SM gauge
group. As such, it can acquire very large mass since there is no gauge
symmetry that forbids it. Thus, a sterile neutrino naturally has very
large mass. There are, however, several ideas that produce naturally
light sterile states \cite{light-st}.

Even if there are light sterile states, complicated patterns of masses
and mixing angles are required in order to accommodate all the
data. In fact, even adding a sterile state does not help in obtaining
a good fit to the data \cite{ster}. In the following we do not address
the four neutrino mixing any further.

%
\section{Models for neutrino masses}
As we saw there are three different kinds of theoretical challenges
imposed by the neutrino data.  Considering separately the solar and
the atmospheric neutrino data, the challenge is to generate neutrino
masses at the right scale.  Combining them, we would like to find ways
to generate the flavor structure, in particular, to explain mass
hierarchy with large mixing. In order to accommodate also the LSND
result, we should find a mechanism that generates light sterile
neutrinos, and construct even more complicated flavor structure.

The simplest way to generate neutrino masses is the seesaw
mechanism. It is particularly attractive when it appears in models
that are well motivated because they address various theoretical
puzzles. Below we describe two models where neutrino masses are
generated as one aspect of a model. (See \cite{review,cy} for more
examples.)
  
%
\subsection{Grand Unified Theories (GUTs)}
There are several supporting evidences for supersymmetric grand
unification.  First, in the minimal supersymmetric SM (MSSM) the three
gauge couplings unify at one point at a scale of
\beq
\Lambda_{\rm GUT}\sim3\times10^{16}\ {\rm GeV}.
\eeq
Second, one generation of SM fermions fits into 
two $SU(5)$ multiplets and into one multiplet for higher rank GUT
groups.  Neutrino physics provide further support for GUT,
particularly, for GUTs that are based on  $SO(10)$ (or higher rank) gauge
group. Next we describe two of the facts that make $SO(10)$ GUTs
attractive from the point of view of neutrino physics.

The 15 degrees of freedom of one SM fermion generation fall into a
${\bf 16}$ of $SO(10)$. The one extra degree of freedom is a SM
singlet. Thus, it can serve as a right handed neutrino field.  Once
$SO(10)$ is broken, this singlet becomes massive. Consequently, the SM
active neutrinos acquire small masses via the seesaw mechanism.
Namely, neutrino masses are predictions of $SO(10)$ GUT theories.

$SO(10)$ also predicts the rough scale of the neutrino masses.  To see
it recall that $SO(10)$ relates the up type quark masses to the Dirac
masses of the neutrinos.  In particular, the Dirac mass of the
heaviest neutrino is of the order of the mass of the top quark. The
Majorana mass of the singlet is expected to be of the order of the GUT
breaking scale, $M_N = \lambda \, \Lambda_{GUT}$ where $\lambda$ is an
unknown dimensionless number.  Then, the seesaw generated mass of the
heaviest active neutrino is of the order of
\beq
m_3\sim {m_t^2 \over M_N}\sim{10^{-3} \,{\rm eV} \over \lambda}\,.
\eeq
For $\lambda \sim O(10^{-1})$ it is the mass scale that was found by
solar neutrinos.  Somewhat smaller $\lambda$ is needed in order to get
the scale that is relevant for atmospheric neutrino
oscillation. Generally, it is fair to say that 
$SO(10)$ based theories predict neutrino masses in
accordance with the experimental findings.

%
\subsection{Flavor physics}
The most puzzling features of the SM fermion parameters are their
smallness and the hierarchies between them.  These features suggest
that there is a more fundamental theory where the hierarchies are
generated in a natural way. Generally, a broken horizontal symmetry is
invoked.  (By horizontal symmetry we refer to a symmetry that
distinguishes between the three SM generations.)

We demonstrate this broken flavor symmetry idea by working in the
framework of a supersymmetric $U(1)_H$ horizontal symmetry
\cite{gn}. We assume that the low energy spectrum consists of the
fields of the MSSM.  Each of the supermultiplets carries a charge
under $U(1)_H$.  The horizontal symmetry is explicitly broken by a
small parameter $\lambda$ to which we attribute a charge $-1$. Then, the
following selection rules apply: Terms in the superpotential that
carry a charge $n\geq0$ under $H$ are suppressed by $O(\lambda^n)$,
while those with $n<0$ are forbidden due to the holomorphy of the
superpotential.

Explicitly, the lepton parameters arise from the Yukawa terms
\beq
Y_{ij}L_i E_j H_d+{Z_{ij}\over M}L_i L_j H_u H_u.
\eeq
Here $L_i$ ($E_i$) are the lepton doublet (singlet) superfields, $H_u$
($H_d$) is the up type (down type) Higgs superfields, $Y_{ij}$ is a
generic complex dimensionless $3\times3$ matrix that gives masses to
the charged leptons, $Z_{ij}$ is a symmetric complex dimensionless
$3\times3$ matrix that gives Majorana masses to the neutrinos and
$M$ is a high energy scale.  The selection rule implies
\beqa
H(L_i)+H(E_j)+H(H_d)=n\quad \Longrightarrow&& \quad
Y_{ij}=O(\lambda^n) \nonumber \\
H(L_i)+H(L_j)+2H(H_u)=m\quad \Longrightarrow && \quad
Z_{ij}=O(\lambda^m)
\eeqa
where we assume that the
horizontal charges of all the superfields are non-negative integers.
We learn that the dimensionless couplings that are naively of order
unity are suppressed by powers of a small parameter.

Next we demonstrate a realization of the neutrino anarchy scenario
using the above framework.  Consider the following 
set of $U(1)_H$ charge assignments
\beqa
&&H(H_u)=0, \quad 
H(H_d)=0, \\
&&H(L_1)=0, \quad 
H(L_2)=0, \quad 
H(L_3)=0, \nonumber \\ 
&&H(E_1)=8, \quad 
H(E_2)=5, \quad 
H(E_3)=3. \nonumber
\eeqa
Then, the lepton mass matrices have the following forms
\beq \label{neuone}
M^\ell\sim\vev{H_d}
\pmatrix{\lambda^8&\lambda^5&\lambda^3\cr
\lambda^8&\lambda^5&\lambda^3\cr\lambda^8&\lambda^5&\lambda^3\cr},
\qquad
M^\nu\sim{\vev{H_u}^2\over M}
\pmatrix{1&1&1\cr1&1&1\cr1&1&1\cr}.
\eeq
We emphasize that the sign ``$\sim$" implies that we
only give the order of magnitude of the various entries; there is an
unknown (complex) coefficient of $O(1)$ in each entry that we do
not write explicitly. Eq. (\ref{neuone}) predicts large mixing angles
\beq
\sin\theta_{23}\sim 1, \qquad
\sin\theta_{12}\sim 1, \qquad
\sin\theta_{13}\sim 1,
\eeq
non-hierarchical neutrino masses
\beq
m^\nu_1\sim{\vev{H_u}^2\over M}, \qquad
m^\nu_2\sim{\vev{H_u}^2\over M}, \qquad
m^\nu_3\sim{\vev{H_u}^2\over M}, 
\eeq
and hierarchical charged lepton masses
\beq
m_e \sim \vev{H_d} \lambda^8, \qquad
m_\mu \sim \vev{H_d} \lambda^5, \qquad
m_\tau \sim \vev{H_d} \lambda^3.
\eeq
We see that this model predicts neutrino masses and mixing angles that
are in accordance with the neutrino anarchy scenario.  When we assume
that both $\vev{H_d}$ and $\vev{H_u}$ are at the weak scale and that
$\lambda \sim 0.2$, this model also reproduces the order of magnitude
of the observed charged lepton masses.

Other, more complicated, flavor structure can also be generated using
a similar framework. See \cite{review,cy} for more details.

%
\section{Conclusions}
Neutrino physics is important since it is a tool for probing unknown
physics at very short distances. Since we know that the SM has to be
extended, it is of great interest to search for such new physics. In
general, new physics at high scale predicts massive neutrinos.
Therefore, there is a strong theoretical motivation to look for
neutrino masses.

In recent years various neutrino oscillation experiments found
strong evidences for neutrino masses and mixing 
\bim
\item 
Atmospheric neutrinos show deviation from the expected ratio between
the fluxes of muon neutrinos and electron neutrinos.  Moreover, the muon
neutrino flux has strong zenith angle dependence. The simplest
interpretation of these results is that there are $\nu_\mu-\nu_\tau$
oscillations.
\item
The electron neutrino flux from the Sun is smaller than the
theoretical expectation. Furthermore, the suppression is energy
dependent. The total neutrino flux, as measured using neutral current
reaction by the SNO experiment, agrees with the theoretical
expectation.  The simplest interpretation of these results is that
there are $\nu_e-\nu_x$ oscillations, where $\nu_x$ can be any
combination of $\nu_\mu$ and $\nu_\tau$.
\item
The K2K long baseline experiments found indications for muon neutrino
disappearance.  Moreover, the same neutrino parameters that account for
the atmospheric neutrino disappearance also explain the K2K data.
\item
The KamLAND experiment found evidences for $\overline\nu_e$
disappearance. The data provide an independent test of the same parameter
space that is probed by solar neutrinos.
\eim

The theoretical expectation and the experimental data fit
comfortably. Yet, there are many unsolved problems associated with
neutrino physics. In the future we expect to have more data and thus
we will be able to learn more about neutrinos, and eventually, about
the short distance new physics.

\section*{Acknowledgments}
I thank Yossi Nir, Subhendu Rakshit, Sourov Roy, Ze'ev Surujon and Jure
Zupan for comments on the manuscript.  Y.G. is supported by
the United States--Israel Binational Science Foundation through Grant
No.~2000133, by a Grant from the G.I.F., the German--Israeli
Foundation for Scientific Research and Development, and by the
Israel Science Foundation under Grant No.~237/01.

  
\end{document}